\documentclass[12pt]{article}
\pdfoutput=1
\usepackage[total={6.5in,8.75in}, top=2.4cm, left=2.4cm]{geometry}
\usepackage{lineno}
\usepackage{graphicx}
\usepackage[round,colon,authoryear]{natbib}

\usepackage{bm}
\usepackage{float}
\usepackage{amsfonts}
\usepackage{verbatim}
\usepackage{soul}
\usepackage{color}
\usepackage{setspace}
\usepackage{rotating}

\linenumbers

\usepackage{indentfirst}  

\usepackage{graphicx}

\floatname{panel}{Panel}
\newfloat{panel}{h}{txt}

\begin{document}

\begin{center}
\Large \textbf{
Integrating
Resource Selection Information with Spatial Capture-Recapture
}
\end{center}

\noindent J. Andrew Royle, U.S. Geological Survey, Patuxent
Wildlife Research Center, Laurel, Maryland, 20708,
\emph{email:} aroyle@usgs.gov \\

\noindent Richard B. Chandler,  U.S. Geological Survey, Patuxent
Wildlife Research Center, Laurel, Maryland, 20708,
\emph{email:} rchandler@usgs.gov \\

\vspace{.2in}

{\bf Running title.} Resource Selection and Spatial Capture-Recapture

\vspace{.2in}

{\bf Word count.} 5670

\vspace{.2in}

{\bf Summary.}

1.
Understanding space usage and resource selection is a primary focus of
many studies of animal populations. Usually, such studies are based on
location data obtained from telemetry,
and resource selection functions (RSF) are used for inference.
Another important focus of wildlife research is estimation and
modeling population size and density.
Recently developed spatial capture-recapture (SCR) models accomplish this objective
using individual encounter history data with auxiliary spatial information
on location of capture. SCR models include 
encounter probability functions that 
are intuitively related to RSFs, but 
to date, no one has extended
SCR models to allow for explicit inference about
space usage and resource selection.

2. In this paper we develop the first statistical framework for
jointly modeling space usage, resource selection, and population
density by integrating SCR data, such as from camera traps, mist-nets, or
conventional catch-traps, with resource selection data from telemetered individuals.
We provide a framework for estimation based on marginal
likelihood, wherein we estimate simultaneously the parameters of the
SCR and RSF models.

3.
Our method leads to increases in precision
for estimating 
population density and parameters of ordinary
SCR models.  Importantly, we also find that SCR models {\it alone} can
estimate parameters of resource selection functions and, as such, SCR
methods can be used as the sole source for studying space-usage;
however, precision will be higher when telemetry data are available.

4. Finally, we find that SCR
models using standard symmetric and stationary encounter probability
models produce biased estimates of density when animal space
usage is related to a landscape covariate. Therefore, it is
important that space usage be taken into consideration, if
possible, in studies focused on estimating density using
capture-recapture methods.

\vspace{.2in}



\section{Introduction}

Spatial capture-recapture (SCR) models are relatively new methods for
inference about population density from capture-recapture data using
auxiliary information about individual capture locations
\citep{efford:2004,borchers_efford:2008, royle_young:2008}.  SCR
models posit that $N$ individuals are located within a region denoted
$\mathcal{S}$. Each individual has a home range or activity area
within which movement occurs during some well-defined time interval,
and the center of the animal's activity has Cartesian coordinates
$\mathbf{s}_i$ for individuals $i=1,\ldots,N$. The population is sampled using $J$
traps with coordinates ${\bf x}_{j}$ for $j=1,\ldots,J$, and encounter
probability is expressed as a function of the distance between trap
location (${\bf x}_{j}$), and individual activity center (${\bf
  s}_{i}$).
While SCR models are a relatively recent innovation, their use is already becoming widespread
\citep{efford_etal:2009ecol, gardner_etal:2010jwm,
  gardner_etal:2010ecol,kery_etal:2010, gopalaswamy_etal:2012,
  foster_harmsen:2012} because they resolve critical problems with
ordinary non-spatial capture-recapture methods such as ill-defined
area sampled and heterogeneity in encounter probability due to the
juxtaposition of individuals with traps
\citep{borchers:2011}. 
Furthermore, unlike traditional capture-recapture methods, SCR models
allow for inference about the processes determining 
spatial variation in population density.

Despite the increasing popularity of SCR models, every
application of them has been based on encounter probability
models, such as the bivariate normal distribution, that imply
symmetric and stationary (invariant to translation) models for home range.
While such simple models might be necessitated in practice by sparse
data, 
home range size and shape are often not well represented by stationary
distributions because animals select
resources 
that 
are unevenly distributed in space. Therefore 
more complex models are needed to relate the capture 
process with the way in which individuals utilize
space. 

In this paper, we extend SCR capture probability models to accommodate
models of space usage or resource selection, by extending them to
include one or more explicit landscape covariates, which the
investigator believes might affect how individual animals use space
within their home range (this is what \citep{johnson:1980} called {\it
  third-order} selection). We do this in a way that is entirely
consistent with the manner in which parameters of classical resource
selection functions (RSF) \citep{manly_etal:2002} or utilization
distributions (UD) \citep{worton:1989, fieberg:2005, fieberg:2007} are
estimated from animal telemetry data.  In fact, we argue that SCR
models and RSF/UD models estimated from telemetry are based on the
same basic underlying model of space usage. The important distinctions
between SCR and RSF studies are that (1) resource selection studies do
not result in estimates of population density and (2) in SCR studies,
encounter of individuals is imperfect (i.e., ``$p<1$'') whereas, with
RSF data obtained by telemetry, encounter is perfect.  With respect to
the latter point, we can think of the RSF and SCR studies as being
exactly equivalent either if we have a dense array of trapping
devices, or if our telemetry apparatus 
samples time or space imperfectly.
A key concept that we must confront in order to unify and integrate
SCR and RSF data is that we need to formulate both models in terms of
a common latent variable so that we can make them consistent with
respect to some underlying space utilization process. As we will
explain, this latent variable is the number of times that an
individual uses a particular region 
of the landscape over some period of time.

The modeling framework we develop here simultaneously resolves three
important 
problems: (1) it generalizes all existing capture probability models
for SCR data to accommodate realistic patterns of space usage that
result in asymmetric and irregular home
ranges; 
(2) it allows estimation of RSF parameters directly from SCR data,
i.e., {\it absent} telemetry data; and (3) it provides the basis for
integrating telemetry data directly into SCR models to improve
estimates of model parameters, including density.  Our model greatly
expands the applied relevance of SCR methods for conservation and
management, and for addressing applied and theoretical questions
related to animal space usage and resource selection.

\section{Spatial Capture-Recapture}

A number of distinct observation models have been proposed for spatial
capture-recapture studies \citep{borchers_efford:2008,
  royle_etal:2009ecol, efford_etal:2009ecol}, including Poisson,
multinomial, and binomial observation models.
Here we focus on the binomial model in which we suppose that the $J$
traps 
are operated for $K$ periods (e.g., nights), and the observations are
individual- and trap-specific counts $y_{ij}$, which are binomial with
sample size $K$ and capture probabilities $p_{ij}$ which depend on
trap locations ${\bf x}_{j}$ and individual activity centers ${\bf
  s}_{i}$ as described subsequently.  The vector of trap-specific
counts for individual $i$, ${\bf y}_{i} = (y_{i1},\ldots,y_{iJ})$ is
its {\it encounter history}.  A standard encounter probability model
\citep{borchers_efford:2008} is the Gaussian model in which
\begin{equation}
\log(p_{ij})= \alpha_{0} + \alpha_{1} d_{ij}^{2}
\label{eq.encounter}
\end{equation}
or, equivalently,
$p_{ij} = \lambda_{0} \exp(-d_{ij}^{2} /(2\sigma^{2}) )$,
where 
$d_{ij}$ is the Euclidean
distance between points ${\bf s}_{i}$ and ${\bf x}_{j}$,
$d_{ij} = \|\textbf{s}_i - \textbf{x}_j\| = \sqrt{(s_{i1}-x_{j1})^2 +
  (s_{i2}-x_{j2})^2}$,
and
 $\alpha_{0} = log(\lambda_{0})$ and $\alpha_{1} = -1/(2\sigma^2)$.
Alternative detection models are used, but all are
functions of Euclidean distance and so we do not consider them
further here.

The primary motivation behind our work is that, in all previous
applications of SCR models, simple encounter probability models based
only on Euclidean distance have been used, with estimation based on
standard likelihood or Bayesian methods. These methods regard the
activity center for each individual $i$, ${\bf s}_{i}$, as latent
variables and remove them from the likelihood either under a model of
``uniformity'' in which ${\bf s} \sim \mbox{Unif}({\cal S})$ where
${\cal S}$ is a spatial region (the ``state-space'' of ${\bf s}$), or
a model in which covariates might affect the spatial distribution of
individuals \citep{borchers_efford:2008}. The state-space ${\cal S}$
defines the potential values for any activity center ${\bf s}$, e.g.,
a polygon defining available habitat or range of the species under
study.

A critical problem with standard SCR models is that the encounter
probability model based on Euclidean distance metric is unaffected by
habitat or landscape structure, and it implies that the space used by
individuals is stationary and symmetric, which may be unreasonable in
many applications.  For example, if the common detection model based
on a bivariate normal probability distribution function is used, then
the implied space usage by {\it all} individuals, no matter their
location in space or local habitat conditions, is symmetric with
circular contours of usage intensity.  Subsequently we provide an
extension of this class of SCR models that accommodates asymmetric,
irregular and spatially heterogeneous models of space usage.  Thus,
``where'' an individual lives on the landscape, and the state of the
surrounding landscape, will determine the character of its usage of
space. In particular, we suggest encounter probability models that
imply irregular, asymmetric and non-stationary home ranges of
individuals and that are sensitive to the local landscape being used
by an individual.

\section{Basic Model of Space Usage}
\label{rsf.sec.rsfmodel}

We develop the model here in terms of a discrete landscape purely for
computational expediency. This formulation will accommodate the vast
majority of actual data sets, as almost all habitat or landscape
structure data comes to us in the form of raster data.  Let ${\bf
  x}_{1},\ldots,{\bf x}_{nG}$ identify the center coordinates of a set
of $nG$ pixels that define a landscape.  In SCR studies, a {\it
  subset} of the coordinates ${\bf x}$ will correspond to trap
locations where we might observe individuals whereas, in telemetry
studies, animals are observable (by telemetry fixes) at potentially
{\it all} coordinates.

Let $z({\bf x})$ denote a covariate measured (or defined) for every
pixel ${\bf x}$. For clarity, we develop the basic ideas here in terms of a
single covariate but, in practice, investigators typically have more
than 1 covariate, which poses no additional problems.
We suppose that a population of individuals wanders around space in
some manner related to the covariate $z({\bf x})$, and their locations
accumulate in pixels by some omnipotent accounting mechanism. We will
define ``use of ${\bf x}$'' to be the event that an individual animal
appeared in some pixel ${\bf x}$.
This is equivalently stated in the
literature in terms of individual having {\it selected} ${\bf  x}$.
 As a biological matter,
use is the outcome of individuals moving around their home range \citep{hooten_etal:2010},
i.e., where an individual is at any point in time is the result of
some movement process. However, to understand space usage, it is not
necessary to entertain explicit models of movement, just to observe
the outcomes, and so we don't elaborate further on what could be
sensible or useful models of movement.

Suppose that an individual  is monitored over some period of time
and a fixed number, say $R$, of use observations are recorded.
Let $n({\bf x})$ be the use frequency of pixel ${\bf x}$ for that individual.
i.e., the number of times that individual used pixel ${\bf x}$
during some period of time.
We assume the
following probability distribution for the $nG \times 1$ vector of use
frequencies:
\[
{\bf n} \sim \mbox{Multinom}(R, {\bm \pi})
\]
where ${\bm \pi}$ is the $nG \times 1$ vector of use probabilities
with elements (for each pixel):
\[
 \pi({\bf x}) = \frac{ \exp( \alpha_{2} z({\bf x}) ) }
   { \sum_{x}    \exp(\alpha_{2} z({\bf x})) }
\]
This is the standard RSF model \citep{manly_etal:2002} used to model
telemetry data.
The parameter $\alpha_2$ is the effect of the
landscape covariate $z({\bf x})$ on the relative probability of
use. Thus, if $\alpha_2$ is positive, the relative probability of use
increases as the value of the covariate increases.
In practice, we don't get to
observe $\{ n({\bf x}) \}$ for all individuals but, instead, only for a small
subset say $i=1,2,\ldots, N_{tel}$,  which we capture and install telemetry devices on.
For the telemetered individuals, we assume they behave according
to the same RSF model as the population as a whole, which might be
justified if individuals are randomly sampled from the population.

We extend this model slightly to make it more realistic spatially and
also consistent with standard SCR models. Let
${\bf s}$ denote the centroid of an individuals home range and let
$d({\bf s},{\bf x}) = ||{\bf x} - {\bf s}||$ be the distance from the home
range center ${\bf s}$ of some individual to pixel ${\bf x}$, and let
$n({\bf x},{\bf s})$ denote the use frequency of pixel ${\bf x}$ for an individual with
activity center ${\bf s}$.
We modify the space usage model to accommodate that space
use will be concentrated around an individual's home range center \citep{johnson_etal:2008,forester_etal:2009}:
\begin{equation}
 \pi({\bf x}|{\bf s})  = \frac{ \exp( -\alpha_{1} d({\bf x},{\bf s})^{2} +\alpha_{2} z({\bf x}) )}
{ \sum_{x} \exp(-\alpha_{1} d({\bf x},{\bf s})^{2} +\alpha_{2} z({\bf x}))}
\label{rsf.eq.rsf}
\end{equation}
where $\alpha_1=1/(2\sigma^2)$ describes the rate at which encounter
probability declines as a function of distance, $d({\bf x},{\bf
  s})$. From ordinary telemetry data, it would be possible to estimate
parameters $\alpha_{1}$, $\alpha_{2}$ and also the activity centers
${\bf s}$ using standard likelihood methods based on the multinomial
likelihood \citep{johnson_etal:2008}.

Note that Eq.~\ref{rsf.eq.rsf} resembles standard encounter models
used in spatial capture-recapture but with an additional covariate
$z({\bf x})$.  The main difference between this observation model and
the standard SCR model is that the model here includes the normalizing
constant $\sum_{x} \exp(-\alpha_{1} d({\bf x},{\bf s})^{2} +\alpha_{2}
z({\bf x}))$, which ensures that the use distribution is a proper
probability density function. Thus we are able to characterize the
probability of encounter in terms of both distance from activity
center and space use.
Note that, under this model for space usage or resource selection, if
there are {\it no} covariates, or if $\alpha_{2} = 0$, then the
probabilities $\pi({\bf x}|{\bf s})$ are directly proportional to the
SCR model for encounter probability.  For example, setting $\alpha_{2}
= 0$, then this implies probability of use for pixel ${\bf x}$ is:
\[
p({\bf x}|{\bf s}) \propto  \exp( -\alpha_{1} d({\bf x},{\bf s})^{2}).
\]
Therefore,
for whatever model we choose for
$p({\bf x},{\bf s})$ in an ordinary SCR model, we can modify the distance
component in the RSF function in Eq. \ref{rsf.eq.rsf} accordingly to
be consistent with that model, by choosing $\pi({\bf x}|{\bf s})$
according to
\[
\pi({\bf x}|{\bf s}) \propto \exp( log(p({\bf x}|{\bf s})) + \alpha_{2} z({\bf x}))
\]

As an illustration of space usage patterns under this model, we
simulated a covariate that represents variation in habitat structure
(Fig. \ref{rsf.fig.habitat}) such as might correspond to habitat
quality.
This was simulated by using a
simple kriging interpolator  of spatial noise.
Space usage patterns for
 8 individuals in this landscape are shown in Fig. \ref{rsf.fig.homeranges},
simulated with $\alpha_{1} = 1/(2\sigma^2)$ with $\sigma = 2$ and the
coefficient on $z({\bf x})$ set to $\alpha_{2} = 1$.
These space usage densities -- ``home ranges'' -- exhibit clear
non-stationarity in response to the structure of the underlying
covariate, and they are distinctly asymmetrical.  We note that if
$\alpha_{2}$ were set to 0, the 8 home ranges shown here would
resemble bivariate normal kernels with $\sigma = 2$.  Another
interesting thing to note is that the activity centers are not
typically located in the pixel of highest use or even the centroid of
usage. That is, the observed ``average'' location is not an unbiased
estimator of ${\bf s}$ under the model in Eq. \ref{rsf.eq.rsf}.

\subsection{Poisson use model}

A natural way to motivate this specific model of space usage is to
assume that individuals make a sequence of random resource selection
decisions so that the outcomes $n({\bf x})$ (for all ${\bf
  x}$) are marginally {\it
  independent} Poisson random variables:
\[
 n({\bf x})|{\bf s} \sim \mbox{Poisson}( \lambda({\bf x}|{\bf s}))
\]
where
\[
 \log(\lambda({\bf x}|{\bf s})) = a_{0} -\alpha_{1} d({\bf x},{\bf s})^{2} +  \alpha_{2} z({\bf x})
\]
 In this case, the number of visits to any particular cell is affected
by the covariate $z({\bf x})$ but has a baseline rate ($\exp(a_{0})$)
related to the amount of movement occurring over some time interval.
This is an equivalent model to the multinomial
model given previously in the sense that, if we condition on the total
sample size $R = \sum_{x} n({\bf x})$, then the vector of use
frequencies $\{ n({\bf x}) \}$ for individual with activity center ${\bf s}$,
has a multinomial distribution with probabilities
\[
 \pi({\bf x}|{\bf s}) = \frac{\lambda({\bf x}|{\bf s})}{\sum_{x}
   \lambda({\bf x}|{\bf s})}
\]
which is the same as Eq. \ref{rsf.eq.rsf} because $a_{0}$ cancels from
the numerator and denominator of the multinomial cell probabilities
and thus this parameter is not relevant to understanding space usage.
Note that if use frequencies are summarized over
$i=1,2,\ldots,N_{tel}$ individuals for each pixel, then a standard
Poisson regression model for the resulting ``quadrat counts'' is
reasonable. This corresponds to ``Design 1'' in
\citet{manly_etal:2002}.

\subsection{Random Thinning}

Suppose our sampling is imperfect so that we only observe a smaller
number of telemetry fixes than actual use frequency, $n({\bf x})$.
We express this ``thinning'' (or sampling) by assuming the observed number of
uses is a binomial random variable based on a sample of size $n({\bf
  x})$:
\[
 m({\bf x}) \sim \mbox{Bin}(n({\bf x}), \phi_{0}).
\]

 Then, the marginal distribution of the new random variable $m$ is
 also Poisson
but with mean
\[
 log(\lambda({\bf x}|{\bf s})) = log(\phi_{0}) + a_{0} -\alpha_{1}
 d({\bf x}|{\bf s})^{2} +  \alpha_{2} z({\bf x}).
\]
 Thus, the space-usage model (RSF) for the
thinned counts $m$ is the same as the space-usage model for the
original variables $n$.  This is because if we remove $n$
from the conditional
 model by summing over its possible values, then the vector of thinned
 use frequencies
${\bf m}$ (i.e., for all pixels) is {\it also}  multinomial with cell probabilities
\[
\pi({\bf x}|{\bf s}) = \frac{\lambda({\bf x}|{\bf s})}{\sum_{x}  \lambda({\bf x}|{\bf s})}
\]
and so the constants $a_{0}$ and $\phi_{0}$
cancel from both the numerator and
denominator. Thus, the underlying RSF model applies to the true
unobserved count frequencies ${\bf n}$ and also those produced
by a random thinning or sampling process, ${\bf m}$.

In summary, if we conduct a telemetry study of $i=1,2\ldots,N_{tel}$
individuals, the observed data are
the $nG \times 1$  vectors of use frequencies ${\bf m}_{i}$ for each individual.
 We declare these data to be
``resource-selection data'' which are typical of the type used to
estimate resource-selection functions (RSFs) \citep{manly_etal:2002}.
In fact, the situation we have described here in which 
we obtain a random sample
of use locations and a complete census of available locations is
referred to as ``Design 2'' by \citep{manly_etal:2002}.

\subsection{Resource Selection in SCR Models}

The key to combing RSF data with SCR data is to work with this
underlying resource utilization process and formulate SCR models in
terms of that process.  Imagine that we have a sampling device, such
as a camera trap, in {\it every} pixel. If the device operates
continually then it is no different from a telemetry instrument.  If
it operates intermittently {\it or} does not expose the entire area of
each pixel then a reasonable model for this imperfect observation is
the ``thinned'' binomial model given above, where $\phi_{0}$
represents the sampling effectiveness of the device.  For data that
arise from SCR studies, the frequency of use for each pixel {\it where}
  a trap is located serves as an intermediate latent variable that we
don't observe. From a design standpoint, the main difference between
SCR studies and telemetry is that, for SCR data, we do not have
sampling devices in all locations (pixels) in the landscape. Rather,
the data are only recorded at a subsample of them, the trap locations,
which we identify by the specific coordinates ${\bf x}_{1},\ldots,{\bf
  x}_{J}$.

So we imagine
that the hypothetical perfect data from a camera trapping study are
the counts $m({\bf x})$ only at the specific trap locations
${\bf x}_{j}$, and for all individuals in the population
$i=1,2,\ldots,N$ 
where $N> N_{tel}$. We denote the individual- and
trap-specific counts by $m_{ij}$ for individual with activity center
${\bf s}_{i}$ and trap location ${\bf x}_{j}$.
 In practice, many (perhaps most)
of the $m_{ij} \equiv m({\bf x}_{j},{\bf s}_{i})$ frequencies will be 0,
corresponding to individuals not captured in certain traps.
We then construct our SCR encounter
probability model based on the
view that these frequencies $m_{ij}$ are {\it latent} variables. In particular,
under the SCR model with binary observations,
 we observe a random variable
$y_{ij} = 1$  if the individual $i$ visited the pixel
containing trap $j$ and was detected.
We imagine that $y_{ij}$ is related to the latent variable $m_{ij}$ being the
event $m_{ij}>0$, as follows:
\[
 y_{ij} \sim \mbox{Bern}(p_{ij})
\]
where
\[
 p_{ij} = \Pr(m_{ij}>0) =  1-\exp(- \lambda({\bf x}_{j}|{\bf s}_{i}))
\]
This is the complementary log-log link relating
$p_{ij}$ to $\log(\lambda_{ij})$, setting
$\lambda_{ij} \equiv \lambda({\bf x}_{j}|{\bf s}_{i})$:
\[
 cloglog(p_{ij}) = log(\lambda_{ij})
\]
where
\[
 \log(\lambda_{ij} ) = \log(\phi_{0}) + a_{0} -\alpha_{1} d({\bf
   x}_{j},{\bf s}_{i})^{2} +  \alpha_{2} z({\bf x}_{j}).
\]
and we collect the constants so that $\alpha_{0} = log(\phi_{0}) +
a_{0}$ is the
 baseline encounter rate which includes
the constant intensity of use by the individual and also the baseline
rate of detection, conditional on use.

\section{The Joint RSF/SCR Likelihood}

To construct the likelihood for SCR data when we have auxiliary
covariates on space usage {\it or} direct information on space usage
from telemetry data, we regard the two samples (SCR and RSF) as
independent of one another. In practice, this might not always be the
case but (1) the telemetry data often come from a previous study;
(2) Or, the individuals are not the same, or cannot be reconciled, even if telemetry study occurs simultaneously;
(3) In cases where we {\it
  can} match some individuals between the two samples, regarding them as
independent should only entail a minor
loss of efficiency
because we are disregarding more precise information on a small number
of activity centers. Moreover, we believe, it is unlikely in practice
to expect the two samples to be completely reconcilable and that the
independence formulation is the most generally realistic.

Regarding the two data sets as being independent, our approach here
is to form the likelihood for each set of observations as a function
of the same underlying parameters and then combine them. In
particular, let ${\cal L}_{scr}(\alpha_{0}, \alpha_{1}, \alpha_{2}, N;
{\bf y}_{scr})$
be the likelihood for the SCR data in terms of the basic encounter
probability parameters and the total (unknown) population size $N$,
and let ${\cal L}_{rsf}(\alpha_{1},\alpha_{2}; {\bf m}_{rsf})$ be the
likelihood for the RSF data based on telemetry which, because the
sample size of such individuals is fixed, does not depend on $N$.
Assuming independence of the two datasets, the
joint likelihood is the product of these two pieces:
\[
{\cal L}_{rsf+scr}(\alpha_{0},\alpha_{1},\alpha_{2},N; {\bf y}_{scr},{\bf
  m}_{rsf})  = {\cal L}_{scr} \times {\cal L}_{rsf}
\]
In what follows, we provide a formulation of each likelihood
component. An  {\bf R} function for obtaining the MLEs of
model parameters is given in Appendix 1.

We adopt the notation $f(\cdot)$ to indicate the probability
distribution of whatever observable quantity is in question. e.g., $f(u)$ is
the marginal distribution of $u$ and $f(u|v)$ is the conditional
distribution of $u$ given $v$, etc. We use $g(\cdot)$ to represent the
probability distribution of latent variables.
The observation model for the SCR data for individual $i$ and trap $j$,
from sampling over $K$ encounter periods, is:
\begin{equation}
f(y_{ij}| {\bf s}_{i}) = \mbox{Bin}(K, p_{ij}({\bm \alpha}))
\label{rsf.mle.eq.cond-on-s}
\end{equation}
where
\[
p_{ij} \equiv  p(d({\bf x}_{j},{\bf s}_{i}), z({\bf x}_{j}); {\bm \alpha})
 = 1-\exp(- \lambda_{ij} )
\]
and
\[
 \lambda_{ij} = \lambda_{0} \exp(- \alpha_{1} d_{ij}^{2} + \alpha_{2}
 z({\bf x}_{j}) )
\]
We emphasize that this is conditional on the latent variables ${\bf
  s}_{i}$ (which appear in the distances $d_{ij}$). For these latent variables we
adopt the standard assumption of uniformity, ${\bf s}_{i} \sim
\mbox{Unif}({\cal S})$ for each individual $i=1,2,\ldots,N$
\citep{royle_young:2008} where ${\cal S}$ is the state-space of the
random variable ${\bf s}$.

The joint distribution of the data for
individual $i$, conditional on ${\bf s}_{i}$, is the product of $J$
binomial terms (i.e., the contributions from each of $J$ traps):
\[
  f({\bf y}_{i} | {\bf s}_{i} , {\bm \alpha}) =
  \prod_{j=1}^{J} \mbox{Bin}(K, p_{ij} ( {\bm \alpha})).
\]
The marginal likelihood \citep{borchers_efford:2008} is
computed by removing ${\bf s}_{i}$, by integration, from the
conditional-on-${\bf s}$ likelihood and regarding the {\it marginal}
distribution of the data as the likelihood. That is, we compute:
\[
  f({\bf y}_{i}|{\bm \alpha}) =
\int_{{\cal S}}  f( {\bf y}_{i} |{\bf s}_{i},{\bm \alpha}) g({\bf s}_{i}) d{\bf s}_{i}
\]
 where, under the uniformity assumption, we have
$g({\bf s}) = 1/||{\cal S}||$.
The joint likelihood for all $N$ individuals,
is the product of $N$ such terms:
\[
{\cal L}_{scr}({\bm \alpha} | {\bf y}_{1},{\bf y}_{2},\ldots, {\bf y}_{N}) = \prod_{i=1}^{N}
f({\bf y}_{i}|{\bm \alpha})
\]
In practice, we don't know $N$ and so we can't just compute the SCR
likelihood in this manner. Instead, we compute the contributions of
the $n$ observed individuals directly as given above, but then we have
to compute the likelihood contribution for the ``all 0'' encounter
history, i.e., that corresponding to unobserved individuals.  The
mechanics of computing that are the same as for an ordinary observed
encounter history, requiring that we integrate a binomial probability
of ${\bf 0}$ over the state-space ${\cal S}$:
\[
\pi_{0} = \Pr({\bf y} = {\bf 0}) = \int_{{\cal S}}
  f({\bf 0} | {\bf s} , {\bm \alpha})  d{\bf s}.
\]
We
then have to deal with the issue that $n$ itself is a random variable,
and that leads to the combinatorial term in front of the likelihood
which involves the total population size $N$. This produces the
conditional-on-$N$ or ``binomial form'' of the likelihood
\citep{borchers_efford:2008,royle:2009}:
\[
\frac{N!}{n! (N-n)!}
\left\{ \prod_{i=1}^{n} f({\bf y}_{i}|{\bm \alpha}) \right\}
\pi_{0}^{N-n}
\]

For the RSF data from the sample of individuals with telemetry devices
we adopt the same basic strategy of describing the
conditional-on-${\bf s}$ likelihood and then computing the marginal
likelihood by averaging over possible values of ${\bf s}$.
We have ${\bf m}_{i}$, the $nG \times 1$ vector of pixel counts for individual $i$,
where these counts are derived from a telemetry study or similar. We
index these elements as $m_{ig}$ for individual $i$ and grid cell $g$,
noting that our index $j$ is reserved only for trap locations, which
are a subset of the $nG$ coordinates ${\bf x}_{1},\ldots, {\bf x}_{nG}$.
The conditional-on-${\bf s}_{i}$ distribution of the telemetry data
from individual $i$ is, omitting the multinomial combinatorial term which does not
depend on parameters,
\[
 f({\bf m}_{i}  | {\bf s}_{i}, {\bm \alpha} ) \propto
\prod_{g=1}^{nG}  \pi({\bf x}_{g}| {\bf s}_{i})^{m_{ig}}
\]
where
\[
 \pi({\bf x}_{g}|{\bf s}_{i})  = \frac{ \exp( -\alpha_{1} d_{ig}^{2} +\alpha_{2} z({\bf x}_{g}) ) }
{ \sum_{g} \exp(-\alpha_{1} d_{ig}^{2} +\alpha_{2} z({\bf x}_{g}))}
\]
The marginal distribution is 
\[
f({\bf m}_{i}|{\bm \alpha}) =    \int_{{\cal S}}  f({\bf m}_{i} |{\bf s}_{i},{\bm \alpha}) g({\bf s}_{i}) d{\bf s}_{i}
\]
and therefore the likelihood for the RSF data is
\[
{\cal L}_{rsf}({\bm \alpha} | {\bf m}_{1},{\bf m}_{2},\ldots, {\bf m}_{Ntel}) = \prod_{i=1}^{Ntel}
f({\bf m}_{i}|{\bm \alpha}).
\]

A key technical aspect of computing these likelihoods is the
evaluation of the 2-dimensional integral over the state-space ${\cal
  S}$, which we approximate (Appendix 1) by a
summation over a fine mesh of points.
We note also that the binomial form of the likelihood here is
expressed in terms of the parameter
$N$, the population size for the landscape defined by ${\cal
  S}$. Given ${\cal S}$, density is computed as $D({\cal S}) =
N/\mbox{area}({\cal S})$. In our simulation study below we report $N$ as the
two are equivalent summaries of the data once ${\cal S}$ is
defined. \citet{borchers_efford:2008} develop a likelihood based on a
further level of marginalization, in which $N$ is removed from the
likelihood by averaging over a  Poisson prior for $N$.

\section{Simulation Analysis}

We carried-out a simulation study using the landscape shown in
Fig. \ref{rsf.fig.habitat}, and based on populations of size $N=100$ and $N=200$
individuals with activity centers distributed uniformly over the
landscape.  This covariate was simulated by generating a field
of spatially correlated noise to emulate a typical patchy habitat
covariate relevant to habitat quality for a species.  We subjected individuals
to sampling over $K=10$ sampling periods, using a $7 \times 7$ array
of trapping devices located on the the integer coordinates $(u*5,v*5)$
for $u,v = 1,2,3,4,5,6,7$. The SCR encounter model was of the form
\[
\mbox{ cloglog}(p_{ij}) = \alpha_{0}  -\frac{1}{2\sigma^{2}} d_{ij}^{2} + \alpha_{2}  z({\bf x}_{j})
\]
with $\alpha_{0} = -2$,  $\sigma =2$ and $\alpha_{2} = 1$.
In the absence of the covariate $z$, this corresponds
to a RSF that is bivariate normal with standard
deviation 2.
These settings yielded an average of about $n=61$ individuals captured for
the $N=100$ case and about $n=123$ for the $N=200$ case. The latter case
represents what we believe is an extremely large sample size based on
our own experience and thus it should serve to gauge the large sample
bias of the likelihood estimator.

In addition to simulating data from this capture-recapture study, we
simulated 2, 4, 8, 12, 16 telemetered individuals
 to assess the
improvement in precision as sample size increases.  For all cases we
observed 20 telemetry fixes {\it per} individual, assuming individuals
were using space according to a RSF model
with the same parameters as those generating the
SCR data. We simulated 500 data sets for each scenario and, for each
data set, we fit 3 models: (i) the SCR only model, in which the
telemetry data were not used; (ii) the integrated SCR/RSF model which
combined all of the data  for jointly estimating model parameters; and
(iii) the RSF only model which just used the telemetry data alone (and
therefore $\alpha_{0}$ and $N$ are not  estimable parameters).
The focus of the simulations was to address the following basic questions:
 (1) how much does
the root mean-squared error (RMSE) of $\hat{N}$ improve as we add or increase the number of
telemetered individuals?  (2) How well does the SCR model do at
estimating the parameter of the RSF with {\it no} telemetry data?  (3)
How much does the precision of the RSF parameter improve if we add SCR
data to the telemetry data?

Results for $N=100$, $N=200$ and $N_{tel}=(2,4,8,12,16)$ are presented
in Table \ref{tab.results1}. We note that the first row of each batch
(labeled ``\mbox{\tt SCR only}'') represent the same estimator and data
configuration. These replicate runs of the SCR-only situation give us
an idea of the inherent MC error in these simulations, which is
roughly about 0.25 and 0.89 on the $N$ scale for the $N=100$ and
$N=200$ cases, respectively.  The mean $N$ for the SCR-only 
estimator across all 5 simulations for $N=100$ was
$\mbox{mean}(\hat{N}) = 99.418$, an empirical bias of $0.6\%$. For
$N=200$, the estimated $N$ across all 5 simulations (5 levels of
$N_{tel}$) was $\mbox{mean}(\hat{N}) = 199.712$, an empirical bias of
about $0.15\%$, within the MC error of the true value of $N=200$.  The
results suggests a very small bias of $< 1\%$ in the MLE of $N$ for
both the  SCR-only and combined SCR/RSF estimators.  In practice, we
expect a small amount of bias in MLEs as likelihood theory only
guarantees asymptotic unbiasedness.


In terms of RMSE for estimating $N$, we see that (Table
\ref{tab.results1}), generally, there is about a 5\% reduction in RMSE
when we have at least 2 telemetered individuals. And, although there
is a lot of MC error in the RMSE quantities, it might be as much as a
10\% reduction as the sample size of captured individuals increases
under the higher $N=200$ setting. This incremental improvement in RMSE
of $\hat{N}$
makes sense because, while the
telemetry provides considerable information about the structural
parameters of the model, it provides no information about mean $p$,
i.e. $\alpha_{0}$, which comes only from the SCR data. Thus estimating
$N$ benefits only slightly from the addition of telemetry data.

The MLE of the RSF parameter $\alpha_{2}$ exhibits negligible or no
bias under {\it both} the SCR only and SCR/RSF estimators. It is
well-estimated from SCR data alone and even better than RSF data alone
(in terms of RMSE) until we have more than 200 or so telemetry
observations.  The biggest improvement from the use of telemetry data
comes in estimating the parameter $\sigma$. We see that $\hat{\sigma}$
is effectively unbiased, and there is a very large improvement in RMSE
of $\hat{\sigma}$, perhaps as much as 50-60\% in some cases, when the
telemetry data are used in the combined estimator (that really doesn't
translate much into improvements in estimating $N$ as we saw
previously).  Improvement due to adding telemetry data diminishes as
the expected sample sizes increases, and so telemetry data does less
to improve the precision of $\hat{\sigma}$ and $\hat{\alpha}_{2}$ for
$N=200$ than for $N=100$. This is because the SCR data along are
informative about both of those parameters.

The results as they concern likelihood estimation of $N$ suggest that
there is not a substantial benefit to having telemetry
data. Estimators ``SCR only'' and ``SCR/RSF'' both appear
approximately unbiased for $N=100$ and $N=200$, and for any sample
size of telemetered individuals. The RMSE is only 5-10\% improved with
the addition of telemetry information.  However, we find that there is
substantial bias in $\hat{N}$ if we use the {\it misspecified} model
that contains no resource selection component. That is if we leave the
covariate $z({\bf x})$ out of the model and incorrectly fit a model
with symmetric and spatially constant encounter model, we see about
20\% bias in the estimates of $N$ in a limited simulation study that
we carried-out (Tab. \ref{tab.bias}). As such, accounting for resource
selection is important, even though, when accounted for, telemetry
data only improves the estimator incrementally.
In addition, we find that the importance of telemetry data is
relatively more important for smaller sample sizes. We carried-out one
simulation study for the $N=100$ case but with lower average encounter
probabilities, setting $\alpha_{0}=-3$. This produces
relatively smaller data sets with  $E[n] = 37$. The results are shown
in Tab. \ref{tab.lowp}. There are some important features evident from
this table. First, as a result of the small samples, the MLE of $N$ is
biased for both SCR only and SCR/RSF estimators although less biased
for the SCR/RSF estimator than for SCR only. The persistent bias in
$\hat{N}$ for both models results from the information about
$\alpha_{0}$ coming only from SCR data, and that estimator itself is
intrinsically biased in small samples.  Conversely, the estimator of
$\alpha_{2}$, the RSF parameter, appears unbiased for all 3 estimators
(SCR only, SCR/RSF and RSF only), as does the estimator of $\sigma$.
We see relatively larger improvements in RMSE (compared with Tab. \ref{tab.results1}) of $\hat{N}$, and those improvements
increase substantially as $N_{tel}$ increases.

\section{Discussion}

How animals use space is a fundamental interest to ecologists, and
important in the conservation and management of many species.
Normally this is done
by telemetry and models referred to as resource selection functions
\citep{manly_etal:2002}.  Conversely, spatial capture-recapture models
have grown in popularity over the last several years
\citep{efford:2004,borchers_efford:2008, royle:2008,
  efford_etal:2009ecol,royle_etal:2009ecol, gardner_etal:2010ecol,
  gardner_etal:2010jwm, kery_etal:2010,
  sollmann_etal:2011,mollet_etal:2012,gopalaswamy_etal:2012}. These,
and indeed, most,
 development and applications of SCR models have focused on density
estimation, not understanding space usage.  However, it is intuitive that space
usage should affect encounter probability and thus it should be highly relevant
to density estimation in SCR applications. Despite this, a description
of the
relationship between encounter
 probability and space usage has not
been developed in the literature on spatial
capture-recapture models.  Essentially all
published applications of SCR models to date have been based on
simplistic encounter
probability models that are symmetric
and do not vary across space. One exception is
\citet{royle_etal:2012ecol} who developed SCR models that use
ecological distance metrics (``least-cost path'')
instead of normal Euclidean distance. Here
we developed an SCR model in terms of a basic underlying model of
space or resource use, that is consistent with existing views of
resource selection functions (RSFs) \citep{manly_etal:2002}.

In developing the SCR model in terms of an underlying model of space
usage, we achieve a number of enormously useful extensions of existing
SCR and RSF methods:
(1) We have shown how to integrate classical RSF data from
telemetry with spatial capture-recapture data based on individual
encounter histories obtained by classical arrays of encounter devices
or traps. This leads to an improvement in our ability to estimate
density, and also an improvement in our ability to estimate parameters
of the RSF function.  Thus, the combined model is both an extension of
standard SCR models and also and extension of standard RSF models. As
many animal population studies have auxiliary telemetry information,
the ability to incorporate such information into SCR studies has
enormous applicability and immediate benefits in many studies.
While adding RSF data to SCR data may increase precision of the MLE of $N$
only incrementally, the effect can be more substantial in sparse data sets
and, generally, RSF produces
relatively huge gains in precision in the MLE of $\sigma$.
(2) We have shown that one can estimate RSF model parameters
directly from SCR data {\it alone}.
While further exploration of this point
is necessary,
it does establish clearly that SCR
models {\it are} explicit models of space usage. Because
capture-recapture studies are, arguably, more widespread than
telemetry studies alone, this greatly broadens the utility and
importance of data from those studies.
(3) It is also now clear
that one of the important parameters of SCR models, that controlling
``home range radius'', can be directly estimated from telemetry data
alone.
The combined RSF+SCR model does yield large improvements in estimation
of $\sigma$. As a practical matter, this suggests we could estimate
$\sigma$ entirely from data extrinsic to the SCR study which might
provide great freedom in the design of SCR studies. For example, traps
could be spaced far enough apart to generate relatively few (even no) spatial recaptures,
but dramatically increase the coverage of the population, i.e., the
observed sample size of captured individuals relative to $N$.
(4) Finally, we found that an
ordinary SCR model with symmetric encounter probability model produces
extremely biased estimates of $N$ when the population of individuals
does exhibit resource selection.  As such, it is important to account
for space usage when important covariates are known to influence
space usage patterns.

Use of telemetry data in capture-recapture studies has been suggested
previously. For example, \citet{white_shenk:2001} and
\citet{ivan:2012} suggested using telemetry data to estimate the
quantity ``probability that an individual is exposed to sampling'' but
their estimator requires that individuals are sampled in proportion to
this unknown quantity, which seems impossible to achieve in many
studies. In addition, they do not directly integrate the telemetry
data with the capture-recapture model so that common parameters are
jointly estimated. In fact, they don't acknowledge shared
parameters of the two models.  \citet{sollmann_etal:inprep} did
recognize this, and used some telemetry data to estimate directly the
parameter $\sigma$ from the bivariate normal SCR model in order to
improve estimates of density. This was an important conceptual
development in the sense that it recognized the relationship between
SCR models and models of space usage, but their model did not include
an explicit resource selection component, and they did not implement a
joint estimation framework.

We developed a formal analysis framework here based on marginal
likelihood \citep{borchers_efford:2008}.
In principle, Bayesian analysis does not pose any unique challenges
for this new class of models although we expect some loss of
computational efficiency due to the increased number of times the
components of the likelihood would need to be evaluated.  We imagine
that some problems would benefit from a Bayesian formulation,
however. For example, using an open population model that allows for
recruitment and survival over time \citep{gardner_etal:2010ecol} is
convenient to develop in the {\bf BUGS} language and incorporating
information on unmarked individuals has been done using Bayesian
formulations of SCR models \citep{chandler_royle:2012,
  sollmann_etal:inprep} but, so far, not likelihood methods.

In our formulation of the joint likelihood for RSF and SCR data, we
assumed the data from  capture-recapture and telemetry studies were
independent of one another. This implies that whether or not an
individual enters into one of the data sets has no effect on whether
it enters into the other data set. We cannot foresee situations in
which violation of this assumption should be problematic or invalidate
the estimator under the independence assumption.  In some cases it
might so happen that some individuals appear in {\it both} the RSF and
SCR data sets. In this case, ignoring that information should entail
only an incremental decrease in precision because a slight bit of
information about an individuals activity center is
disregarded. Heuristically, an SCR observation (encounter in a trap)
is like one additional telemetry observation, and so the
misspecification (independence)
regards the
two pieces of information as having separate activity centers.
 Our model pretends that we don't know anything
about the telemetered individuals in terms of their encounter history
in traps.  In principle it shouldn't be difficult to admit a formal
reconciliation of individuals between the two lists. In that case, we
just combine the two conditional likelihoods before we integrate ${\bf
  s}$ from the conditional likelihood. This would be almost trivial to
do if {\it all} individuals were reconcilable (or none as in the case
we have covered here) but, in general , we think you will always have
an intermediate case -- i.e., either none will be or at most a subset
of telemetered individuals will be known. More likely you have variations of ``well, that
guy looks telemetered but we don't know which guy it is....hmmm'' and
that case, basically a type of marking uncertainty or
misclassification, is clearly more difficult to deal with.

We conclude that the key benefit of our combined SCR/RSF model is its
ability integrate realistic patterns of space usage directly into SCR
models and avoid extreme bias in estimating $N$ and, secondarily, we
are able to obtain RSF information from SCR alone.  Therefore, our new
class of integrated SCR/RSF models allows investigators to model how
the landscape and habitat influence movement and space usage of
individuals around their home range, using non-invasively collected
capture-recapture data or capture-recapture data augmented with
telemetry data.  This should improve our ability to understand, and
study, aspects of space usage and it might, ultimately, aid in
addressing conservation-related problems such as reserve or corridor
design. And, it should greatly expand the relevance and utility of
spatial capture-recapture beyond simply its use for density
estimation.

\section*{Acknowledgments}

\bibliography{RSFmanuscript-v2.bbl}

\section*{Appendix 1: {\bf R} script for obtaining MLEs under the SCR+RSF model}

{\small
\begin{verbatim}
### before running this code, put the functions at the end of this script
### into your R workspace
###


## the following block of code makes up a covariate as a spatially correlated
## noise field, with an exponential spatial correlation function
set.seed(1234)
gr<-expand.grid(1:40,1:40)
Dmat<-as.matrix(dist(gr))
V<-exp(-Dmat/5)
z<-t(chol(V))%*%rnorm(1600)
spatial.plot(gr,z)


###
### Set some parameter values
###
alpha0 <- -2
sigma<- 2
beta<- 1
Ntel<-4      # number of individuals with telemeters
nsim<-100
Nfixes<-20   # number of telemetry fixes per individual
N<- 100      # population size


# simulate activity centers of all N individuals
Sid<- sample(1:1600,N,replace=TRUE)
# and coordinates
S<-gr[Sid,]
# now draw centers of telemetered individuals
# have to draw telemetry guys interior or else make up more landscape --
# can't have truncated telemetry obs

poss.tel<- S[,1]>5 & S[,1]<35 & S[,2]>5 & S[,2]<35
tel.guys<-sample(Sid[poss.tel],Ntel)
sid<-tel.guys
stel<-gr[sid,]

# make up matrix to store RSF data
n<-matrix(NA,nrow=Ntel,ncol=1600)

# for each telemetered guy simulate a number of fixes.
# note that n = 0 for most of the landscape
par(mfrow=c(3,3))
lammat<-matrix(NA,nrow=Ntel,ncol=1600)
for(i in 1:Ntel){
   d<- Dmat[sid[i],]
   lam<- exp(1 - (1/(2*sigma*sigma))*d*d + beta* z)
n[i,]<-rmultinom(1,Nfixes,lam/sum(lam))
   par(mar=c(3,3,3,6))
   lammat[i,]<-lam
   img<- matrix(lam,nrow=40,ncol=40,byrow=FALSE)
   image(1:40,1:40,rot(img),col=terrain.colors(10))
}

## now lets simulate some SCR data on a bunch of guys:

# make a trap array
X<-  cbind(  sort(rep( seq(5,35,5),7)), rep( seq(5,35,5),7))
ntraps<-nrow(X)
raster.point<-rep(NA,nrow(X))
for(j in 1:nrow(X)){  # which piont in the raster is the trap? must be raster points
 raster.point[j]<- (1:1600)[ (X[j,1]==gr[,1]) & (X[j,2] == gr[,2])]
}
points(X,pch=20,cex=2)

D<- e2dist(S,X)  ## N x ntraps
Zmat<- matrix(z[raster.point],nrow=N,ncol=ntraps,byrow=TRUE) # note make dims the same
loglam<-   alpha0  -(1/(2*sigma*sigma))*D*D + beta*Zmat
p<- 1-exp(-exp(loglam))

## Now simulate SCR data

K<- 10
y<-matrix(NA,nrow=N,ncol=ntraps)
for(i in 1:N){
y[i,]<- rbinom(ntraps,K,p[i,])
}

cap<-apply(y,1,sum)>0

y<-y[cap,]
gr<-as.matrix(gr)
sbar<- (n%*%gr)/as.vector(n%*%rep(1,nrow(gr)))

# Basic SCR model with RSF covariate at trap locations.
tmp1<-nlm(intlik3rsf.v2,c(-3,log(3),1,0),y=y,K=K,X=X,ztrap=z[raster.point],G=gr)

# use telemetry data and activity centers for those are marginalized out of the likelihood
tmp2<-nlm(intlik3rsf.v2,c(-3,log(3),1,0),y=y,K=K,X=X,ztrap=z[raster.point],G=gr,ntel=n,zall=as.vector(z))

# use mean "s" instead of estimating it
tmp3<-nlm(intlik3rsf.v2,c(-3,log(3),1,0),y=y,K=K,X=X,ztrap=z[raster.point],G=gr,ntel=n,zall=as.vector(z),stel=sbar)

# no SCR data, s is random. Here there are 2 extra parameters that are not estimated: start[1] and start[4]
tmp4<-nlm(intlik3rsf.v2,c(-3,log(3),1,0),y=NULL,K=K,X=X,ztrap=z[raster.point],G=gr,ntel=n,zall=as.vector(z))

# Fits SCR model with isotropic Gaussian encounter model
tmp5<- nlm(intlik3rsf.v2,c(-3,log(3),1,0),y=y,K=K,X=X,ztrap=rep(0,ntraps),G=gr)


###
### put all the functions below this line into your R workspace
###

 spatial.plot<-
function(x,y){
 nc<-as.numeric(cut(y,20))
 plot(x,pch=" ")
 points(x,pch=20,col=topo.colors(20)[nc],cex=2)
 ###image.scale(y,col=topo.colors(20))
}

### This is the likelihood function
### It computes several versions of the likelihood depending on the arguments specified
### see the 5 examples above

intlik3rsf.v2 <-function(start=NULL,y=y,K=NULL,X=traplocs,ztrap,G,ntel=NULL,zall=NULL,stel=NULL){
# start = vector of length 5 = starting values
# y = nind x ntraps encounter matrix
# K = how many samples?
# X = trap locations
# ztrap = covariate value at trap locations
# zall = all covariate values for all nG pixels
# ntel = nguys x nG matrix of telemetry fixes in each nG pixels
# stel = home range center of telemetered individuals, IF you wish to estimate it. Not necessary

nG<-nrow(G)
D<- e2dist(X,G)

alpha0<-start[1]
sigma<- exp(start[2])
alpha2<- start[3]
n0<-    exp(start[4])
a0<- 1

if(!is.null(y)){
loglam<-   alpha0  -(1/(2*sigma*sigma))*D*D + alpha2*ztrap  # ztrap recycled over nG
probcap<- 1-exp(-exp(loglam))
#probcap<- (exp(theta0)/(1+exp(theta0)))*exp(-theta1*D*D)
Pm<-matrix(NA,nrow=nrow(probcap),ncol=ncol(probcap))
ymat<-y
ymat<-rbind(y,rep(0,ncol(y)))
lik.marg<-rep(NA,nrow(ymat))
for(i in 1:nrow(ymat)){
Pm[1:length(Pm)]<- (dbinom(rep(ymat[i,],nG),rep(K,nG),probcap[1:length(Pm)],log=TRUE))
lik.cond<- exp(colSums(Pm))
lik.marg[i]<- sum( lik.cond*(1/nG) )
}
nv<-c(rep(1,length(lik.marg)-1),n0)
part1<- lgamma(nrow(y)+n0+1) - lgamma(n0+1)
part2<- sum(nv*log(lik.marg))
out<-  -1*(part1+ part2)
}
else{
out<-0
}

if(!is.null(ntel) & !is.null(stel) ){

# this is a tough calculation here
D2<-  e2dist(stel,G)^2
# lam is now nG x nG!
lam<- t(exp(a0 - (1/(2*sigma*sigma))*t(D2)+ alpha2*zall))  # recycle zall over all ntel guys
denom<-rowSums(lam)
probs<- lam/denom  # each column is the probs for a guy at column [j]

tel.loglik<-  -1*sum(  ntel*log(probs) )

out<- out  + tel.loglik
}

if(!is.null(ntel) & is.null(stel) ){

# this is a tough calculation here
D2<-  e2dist(G,G)^2
# lam is now nG x nG!
lam<- t(exp(a0 - (1/(2*sigma*sigma))*t(D2)+ alpha2*zall))  # recycle zall over all ntel guys
denom<-rowSums(lam)
probs<- t(lam/denom)  # each column is the probs for a guy at column [j]
marg<- as.vector(rowSums(exp(ntel%*%log(probs))/nG ))

tel.loglik<- -1*sum(log(marg))

out<- out  + tel.loglik
}

out
}

\end{verbatim}
}

\clearpage
\newpage

\begin{sidewaystable}[h!]
\centering
\caption{Mean and RMSE of sampling distribution of the MLE of $N$ and
  other model parameters under a model of resource selection using
  only SCR data, SCR combined with RSF data on $N_{tel}$ individuals,
  and with RSF only data on $N_{tel}$ individuals. Simulations results
  are based on 500 Monte Carlo simulations of populations containing
  $N=100$ or $N=200$ individuals. The true parameter values were
  $\alpha_{2} = 1$ and $\sigma = 2$.
}
\begin{tabular}{ccccccccccccc}
\hline \hline
Estimator &  \multicolumn{6}{c}{N=100} & \multicolumn{6}{c}{N=200} \\
\hline
$N_{tel}=2$       &   $\hat{N}$ & RMSE & $\hat{\alpha}_{2}$& RMSE &
$\hat{\sigma}$ & RMSE & $\hat{N}$ & RMSE & $\hat{\alpha}_{2}$ & RMSE &
$\hat{\sigma}$ & RMSE \\ \hline
SCR only: &  99.73 & 9.97 & 0.99 & 0.14 & 2.00&  0.124& 198.85&14.24&0.99&   0.10& 2.00& 0.091 \\
SCR/RSF:  &  99.94 & 9.54 & 0.99 & 0.12 & 2.00&  0.097& 199.37&12.80&0.99&   0.09& 2.00& 0.078 \\
RSF only  &   --   & --   & 1.03 & 0.33 & 2.00&  0.160&   --  &   --&1.04&   0.33& 1.99& 0.169 \\
$N_{tel}=4$&        &      &      &      &     &       &       &      &    &       &     &      \\
SCR only  &  99.10 & 9.83 & 0.99 & 0.13 & 2.00&  0.127& 200.06& 15.34&1.00&   0.09& 2.00& 0.092\\
SCR/RSF   &  99.17 & 9.47 & 0.99 & 0.11 & 2.00&  0.086& 200.25& 14.36&1.00&   0.08& 2.01& 0.073\\
RSF only  &   --   &  --  & 0.98 & 0.22 & 2.00&  0.119&   --  &  --  &1.02&   0.21& 2.01& 0.122\\
$N_{tel}=8$&        &      &      &      &     &       &       &       &    &        &    &      \\
SCR only  &  99.59 & 10.00& 1.00 & 0.13 & 2.00&  0.130& 200.85& 14.06&1.00&   0.09& 2.00&   0.087\\
SCR/RSF   &  98.90 & 10.02&  0.99&  0.10& 2.00&  0.071& 200.29& 13.98&1.00&   0.08& 2.00&   0.061\\
RSF only  &   --   & --   & 0.98 & 0.16 & 2.01&  0.084&   --  &  --  &0.99&   0.16& 2.00&   0.084\\
$N_{tel}=12$&       &      &      &     &     &       &        &      &      &        &   &        \\
SCR only  &  99.44 &10.73 & 0.98 & 0.13 & 2.02&  0.128& 198.76& 14.47& 0.99&   0.10& 2.00&   0.091\\
SCR/RSF   &  99.96 &10.26 & 1.00 & 0.09 & 2.00&  0.059& 198.72& 14.14& 1.00&   0.08& 2.00&   0.054\\
RSF only  &   --   & --   & 1.01 & 0.12 & 2.00&  0.069&   --  &  --  & 1.01&   0.13& 2.00&   0.069\\
$N_{tel}=16$&        &      &      &     &     &       &       &      &     &       &       &      \\
SCR only  &  99.23 &10.74 & 0.99 & 0.14 & 2.00&  0.128& 200.04& 14.09&0.99&   0.10& 2.01&   0.088  \\
SCR/RSF   &  99.20 & 9.79 & 1.00 & 0.09 & 1.99&  0.057& 200.25& 13.40&1.00&   0.07& 2.00&   0.047 \\
RSF only  &   --   & --   & 1.00 & 0.10 & 1.99&  0.061&   --  &  --  &1.00&   0.11& 2.00&   0.055 \\
\end{tabular}
\label{tab.results1}
\end{sidewaystable}

\begin{table}
\centering
\caption{Expected value of $\hat{N}$ and $\hat{\sigma}$
 for truth
  $N=200$ and $\sigma=2$ under a model of resource selection with a
  single covariate, when the
encounter probability model is misspecified by a symmetric and
constant model assuming no resource selection; column ``bias'' is {\it
  percent} bias.
}
\begin{tabular}{cccccccc}
\hline \hline
    &   $E[\hat{N}]$ & bias & RMSE & $E[\hat{\sigma}]$ & RMSE \\  \hline
n=2 &     161.48 &-19.2 & 39.98  & 1.84 &  0.180  \\
n=4 &     161.32 &-19.3 &40.00  & 1.83 &  0.191 \\
n=8 &     161.46 &-19.3 &40.06  & 1.84 &  0.184 \\
n=12 &    162.40 &-18.8  &38.95  & 1.84 &  0.185\\
n=16 &    160.93 &-19.5 &40.44  & 1.84 &  0.190 \\
\end{tabular}
\label{tab.bias}
\end{table}

\begin{table}
\centering
\caption{Mean and RMSE of the sampling distribution of the MLE for
  model parameters for the $N=100$ and ``low $p$'' case. For each of
  500 simulated data sets, a model was fit using the SCR likelihood
  only, the joint SCR/RSF likelihood, and the RSF likelihood only. For
  the latter, the parameter $N$ is not statistically identifiable. }
\begin{tabular}{ccccccc}
\hline \hline
Estimator & $E[\hat{N}]$ & RMSE & $E[\hat{\alpha}_{2}]$ & RMSE &
$E[\hat{\sigma}]$ & RMSE \\
 \hline
$N_{tel} = 2$ &      &      &       &      &    &      \\
SCR only     &103.85& 22.88&   1.00& 0.19& 2.02& 0.261 \\
SCR/RSF      &102.90& 20.98&   1.00& 0.17& 2.00& 0.136\\
RSF only     &--    & --   &  1.02 & 0.30& 1.99& 0.163\\
$N_{tel}=4$   &      &      &       &     &     &        \\
SCR only     &105.65& 26.52&   1.01& 0.20& 2.01& 0.258 \\
SCR/RSF      &103.55& 22.92&   1.01& 0.14& 2.00& 0.104\\
RSF only     & --   & --   &  1.01 & 0.21& 1.99& 0.114\\
$N_{tel}=8$   &      &      &       &     &     &       \\
SCR only     &107.41&  45.05&   0.99& 0.19& 2.01& 0.254 \\
SCR/RSF      &104.28&  22.13&   1.00& 0.12& 2.00& 0.076\\
RSF only     &--    & --    &  1.01& 0.15& 1.99& 0.081\\
$N_{tel}=12$  &      &       &      &     &      &      \\
SCR only     &106.35&  27.32& 0.99& 0.19 & 2.00& 0.255\\
SCR/RSF      &104.11&  21.81& 1.00& 0.10 & 2.00& 0.063\\
RSF only     & --   &  --   & 1.01& 0.12 & 2.00& 0.065\\
$N_{tel}=16$  &      &       &     &      &     &       \\
SCR only     &104.05&  31.41& 0.99& 0.19 & 2.02& 0.252\\
SCR/RSF      &101.98&  20.78& 1.00& 0.09 & 2.00& 0.055\\
RSF only     & --   &  --   & 1.00& 0.10 & 2.00& 0.056\\
\end{tabular}
\label{tab.lowp}
\end{table}

\clearpage

\newpage

{\flushleft FIGURE CAPTIONS }

\vspace{.2in}

{\flushleft \bf
Figure 1:}
A typical habitat covariate reflecting habitat quality or
  hypothetical utility of the landscape to a species under study. Home range centers for 8 individuals are
shown with black dots.

\vspace{.2in}

\begin{figure}
\centering
\includegraphics[width=3.5in,height=3.25in]{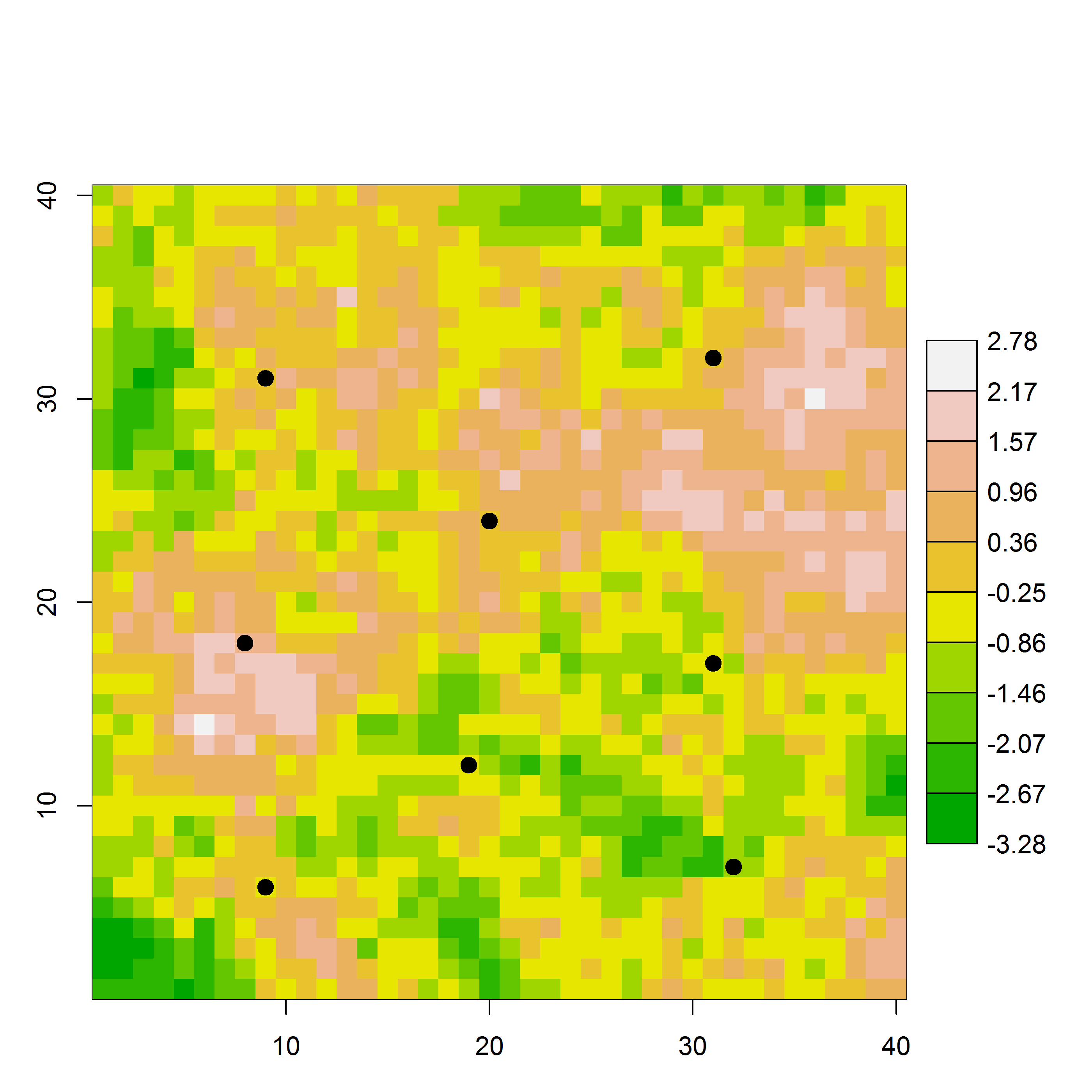}
\caption{A typical habitat covariate reflecting habitat quality or
  hypothetical utility of the landscape to a species under study. Home range centers for 8 individuals are
shown with black dots.}
\label{rsf.fig.habitat}
\end{figure}

\begin{figure}
\centering
\includegraphics[width=3.5in,height=3.5in]{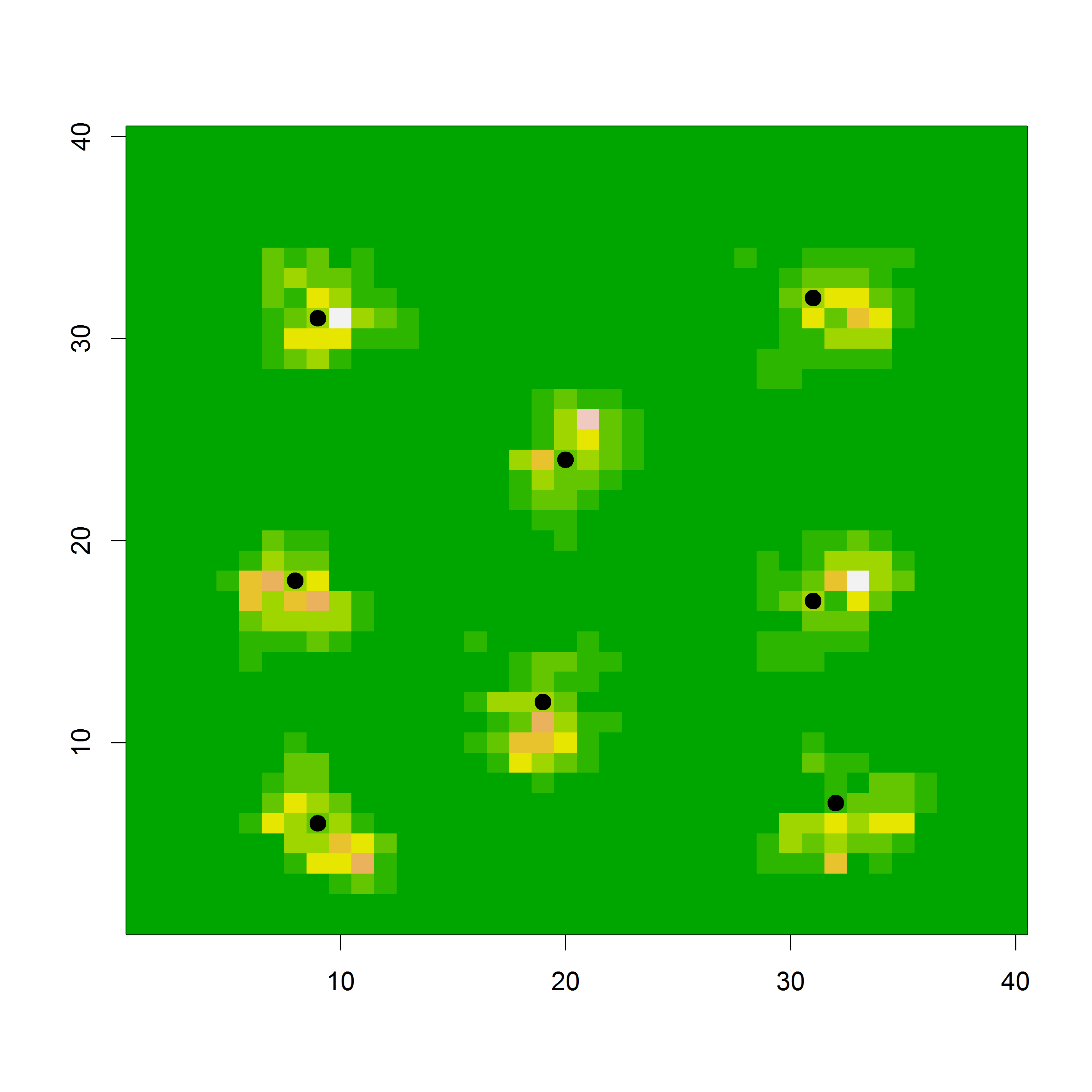}
\caption{Space usage patterns of 8 individuals under a space usage
  model that contains a single covariate (shown in
  Fig. \ref{rsf.fig.habitat}). Plotted value is the multinomial
  probability $\pi_{ij}$ for pixel $j$ under the model in Eq. \ref{rsf.eq.rsf}.
}
\label{rsf.fig.homeranges}
\end{figure}

{\flushleft \bf Figure 2:}
Space usage patterns of 8 individuals under a space usage
  model that contains a single covariate (shown in
  Fig. \ref{rsf.fig.habitat}). Plotted value is the multinomial
  probability $\pi_{ij}$ for pixel $j$ under the model in Eq. \ref{rsf.eq.rsf}.

\newpage

\clearpage
\newpage

\end{document}